\documentclass[conference]{IEEEtran}

\IEEEoverridecommandlockouts %To add sponsors and index terms

\usepackage{graphicx}
\usepackage{multirow}
\usepackage[table,xcdraw]{xcolor}
\usepackage{epstopdf}
\usepackage{amsfonts}
\usepackage{amsmath,amssymb}
\usepackage{color,soul}
\usepackage{color}
\usepackage{verbatim}
\usepackage{comment}
\usepackage{algorithm}
\usepackage[table,xcdraw]{xcolor}
\usepackage{varioref}
\usepackage{gensymb}
\usepackage{comment}
\usepackage{xfrac}
\usepackage{soul}
\usepackage{color}
\usepackage{xcolor}
\usepackage{cite}
\usepackage{ctable}

\usepackage{caption}
\usepackage{subcaption}
\usepackage{cleveref}

\makeatletter

\begin{document}
\title{Multivariate Prediction Intervals for Photovoltaic Power Generation}

\author{%
	
	\IEEEauthorblockN{Faranak~Golestaneh,~\IEEEmembership{Student Member,~IEEE,} \\ }
	\IEEEauthorblockA{School of Electrical and Electronic Engineering\\
		Nanyang Technological University\\
		Singapore\\
		faranak001@e.ntu.edu.sg}
	\and
	\IEEEauthorblockN{Hoay~Beng~Gooi,~\IEEEmembership{Senior Member,~IEEE,} \\ }
	\IEEEauthorblockA{School of Electrical and Electronic Engineering\\
		Nanyang Technological University\\
		Singapore\\
		EHBGOOI@ntu.edu.sg}
	  \thanks{This work was supported by the Energy Innovation Programme Office (EIPO) through the National Research Foundation (NRF) and Singapore Economic Development Board (EDB).}
}

\maketitle
\pagestyle{plain}
%\vspace{-1.5em}
%ABSTRACT
\begin{abstract}
The current literature in probabilistic forecasting
is focused on quantifying the uncertainty of each random
variable individually. This leads to the failure in informing about
interdependence structure of uncertainty at different locations
and/or different lead times. When there is a positive or negative
association between a number of random variables, the prediction
regions for them should be reflected by multivariate or joint
uncertainty sets. The existing literature is very primitive in
the area of multivariate uncertainty sets modeling. In this paper, uncertainty regions are generated in the form of multivariate prediction intervals. We will examine the performance of Gaussian and R-Vine copulas in characterizing the correlated behavior of PV power generations at successive lead-times. Copulas are compared based on goodness-of-fit metrics as well as skill scores.  A framework is elaborated to generate multivariate prediction intervals out of the scenarios generated from Gaussian and R-vine multivariate densities. The resultant multivariate prediction intervals are evaluated based on their calibration and sharpness. The approaches are tested on a real-world dataset including PV power measurements and weather forecasts. This paper provides a series of useful analyses and comparative results for multivariate uncertainty modeling of PV power that can serve as a basis for  future works in the area.\\
\end{abstract}
\vspace{-.5em}
%INDEX TERMS

\vspace{-0.4em}
\section{Introduction}

Deployment of Photovoltaic (PV) energy as one of the highly intermittent types of energies is accelerating worldwide.  However, high integration of PV power technologies can endanger reliability of power grids unless their variable and uncertain nature is actively managed. PV energy forecasting is a vital decision-aiding tool for either balancing authorities or market participants~\cite{flagship2012solar}.

The fact is that there is always a non-negligible level of uncertainty attached to forecasts, meaning that forecasts are
always precisely wrong. Therefore, decision-makers need to have an estimation of the uncertainty while
this cannot be provided by conventional point forecasts. In recent years, there has been a transition from point forecasts to probabilistic forecasting. Probabilistic forecasts provide quantitative information on the uncertainty associated
with each  forecast for the future. Among very few published works in PV power point forecasting one can refer to~\cite{golestaneh2016very}. In~\cite{golestaneh2016very}, the marginal predictive densities of PV power are developed for each lead-time individually, neglecting the high temporal correlations  observed in PV power generations.

To describe PV generations at a number of locations and lead-times, a framework is proposed in~\cite{golestaneh2016generation} to produce joint or multivariate
distributions. In~\cite{golestaneh2016generation}, a Gaussian distribution is formulated based on the  Gaussian copula to describe temporal/spatio-temporal dependences in PV generation. 

While Gaussian copula assumption is the most straightforward one, it might not be accurate in general. For example, PV power is a double-bounded variable limited to 0 and the maximum capacity of PV installation, while the Gaussian distribution describes variables as unlimited with the tails going towards infinity. More investigations are required to explore  the relevance of various copulas to characterize intermittent behavior of  PV power. The problem is that, the number of standard copulas in high dimensions are limited and their inflexible structure  makes forming them very challenging.  Vine copulas have introduced an interesting solution for this problem by breaking the complex patterns of dependence into cascade of pair-copulas. This allows to benefit from the rich variety of bivariate copulas already available as building blocks~\cite{czado2010pair}. Among very few studies available on applications of vine copulas in energy sector,\cite{bessa2016quality} can be mentioned where C-Vine and D-Vine copulas are used to model multivariate dependence of wind power.

%For example, a Gaussian copula model has zero tail dependency. This implies that there is asymptotically zero probability that random  variables are in their extremes unless their Pearson correlation coefficient is unity~\cite{hu2006dependence}. 

%More investigations are required to explore  the relevance of various multivariate densities to characterize intermittent behavior of  PV generation. The rival approaches (copulas) should be examined on their goodness-of-fit as well as skill scores.

%Because in vine copula modeling, multidimensional dependencies are factorized into pair-copulas, Vine copulas add much more flexibility into multidimensional dependency modeling.

 %It is claimed that the vine copula can be more efficient in cases where different pairs of the dependent variables have different dependency structures~\cite{wu2015versatile}. Vine copulas leverage from bivariate copulas and enable extensions to arbitrary dimensions. The number of available parametric bivariate copulas are much more than the available parametric multidimensional copulas. This extends the number of copula options for parametric modeling of high dimensional dependencies~\cite{czado2010pair}. 

Multivariate densities are often represented by 
scenarios, also referred to as trajectories or ensembles. Multivariate scenarios are generated by random draw from the multivariate predictive distributions. However, for
classes of decision-making problems based on robust, interval and
chance-constrained optimization, it is required to represent uncertainties in the form of multivariate uncertainty regions. Prediction intervals can be considered as one of the most straightforward uncertainty representations. Multivariate Prediction intervals (MPIs) provide
information about the probability of having the observed random
variable in consecutive times fully inside the multivariate region. MPIs is still at very stage in energy sector. There are only few attempts made so far to produce
MPIs for energy related variables. A method called adjusted Intervals is used in~\cite{li2011simultaneous} to generate MPIs from Gaussian multivariate scenarios for wind power.

%The scoring rules examine the ability of forecasts in imitating the real stochastic process of interest.
 To the best of our knowledge there is  no published  work in the literature which reports and discusses MPIs of PV power. This paper offers two important contributions in the area of probabilistic forecasting of PV power in a multivariate context. First, it elaborates  a framework to generate MPIs based on R-Vine and Gaussian copulas. R-Vine copula structure provides more flexibility comparing to C-Vine or D-Vine which have more restricted frameworks~\cite{kramer2011introduction}. Second, it compares the relevance and the quality of Gaussian and R-Vine copulas in terms of goodness-of-fit and skill scores for multivariate characterization of PV power generation. The objective is to investigate  which copula presents better goodness-of-fit in describing the correlated PV power generation in successive hours. Then, we  examine whether a higher goodness-of-fit directly lead to more skilled multivariate scenarios and MPIs. The evaluation metrics covered in this work are corrected Akaike's Information Criterion (AIC), the Bayesian Information Criterion (BIC), log-likelihood,  energy and variogram-based scores. MPIs also are evaluated based on their  volume as representative of sharpness and calibration. A real-world dataset including Numerical Weather Predictions (NWPs) and PV power measurements is used for verifications. 
\vspace{-0.6em}
\section{Methodology}

 Let  \textbf{P} be the multivariate PV power generation of dimension $D$ where \textbf{X} is given by   $\textbf{P}_t=\{P_{t+k_1},...,P_{t+k_D}\}$ with $k_i \:\forall i$ as the forecast horizons and $t$ as the day index. To simplify the notation, hereafter $ \textbf{P}_t $ is denoted as $\textbf{P}_t=\{P_{t,1},...,P_{t,D}\}$.
 
 Let $\hat{ \textbf{F}}_t $ be  the predictive multivariate distribution conditional on information up to time $t$ as an estimation of $ \textbf{F}_t $. Denote $\hat{F}_{t,d} \: (d=1,..., D) $ as predictive marginal distributions for forecast horizons individually. Then, according to Sklar's theorem~\cite{sklar1959fonctions}, $\hat{ \textbf{F}}_t $ can be written as
 \begin{equation}
 \label{Eq:coplas}
 \hat{\textbf{F}}_t(\textbf{P}_t)=C(\hat{F}(P_{t,1}),  \hat{F}(P_{t,2}) , ...,  \hat{F}(P_{t,D}))\\
 \end{equation}
 with $C$ as the copula function. $C$ can be considered as the distribution function of a $D$-dimensional random variable with uniformly distributed marginals $U(0,1)$~\cite{brechmann2013modeling}. Corresponding densities will be denoted by a small letter $c$.

  Sklar's theorem allows to model  the multivariate dependence  in terms of the copula separately from the marginal distributions and then link them together to find the multivariate density. 
 
 In~\cite{golestaneh2016generation}, in detail, it is explained how to model multivariate densities and  scenarios using Gaussian copula and marginal densities. Therefore,  here we will elaborate modeling multivariate densities based on R-Vine copulas only. 
\subsection{R-Vine Copula}
\label{subsection:Vine}
In vine copula modeling,  a multivariate density function of dimension $ D $ is constructed using $ D(D-1)/2 $ bivariate copulas~\cite{schepsmeier2013goodness} and a nested set of trees. To demonstrate how pair-copula constructions is defined, here an example for a three dimensional variable $\textbf{P}=(P_1,P_2,P_3)~$ is given
\begin{multline}
\label{Eq:RVineDim3}
{\textbf{f}}(P_{1}, P_{2}, P_{3})=f_{3|12}(P_3|P_1,P_2)f_{2|1}(P_2|P_1)f_1(P_1)
\end{multline}
From \eqref{Eq:coplas} we get,
\begin{multline}
\label{Eq:RVineDim3B}
f_{2|1}(P_2|P_1)= c_{12}(F_1(P_1),F_2(P_2)){f}_2(P_2)\\
f_{3|12}(P_3|P_1,P_2)=c_{13|2}(F_{1|2}(P_1|P_2),F_{3|2}(P_3|P_2))f_{3|2}(P_3|P_1)\\
f_{3|2}(P_3|P_1)=c_{23}(F_2(P_2),F_3(P_3)){f}_3(P_3)
\end{multline}
By substituting \eqref{Eq:RVineDim3B} in \eqref{Eq:RVineDim3},
\begin{multline}
\label{Eq:RVineDim3C}
{f}_3(P_3) {f}_2(P_2) {f}_1(P_1) \\
\times c_{12}(F_1(P_1),F_2(P_2))c_{23}(F_2(P_2),F_3(P_3))\\
c_{13|2}(F_{1|2}(P_1|P_2),F_{3|2}(P_3|P_2))
\end{multline}
with small letter $f_i\:(\forall i)$ representing corresponding densities of $F_i\:(\forall i)$.

Similarly, the pair-copula construction in $D$ dimensions is given by~\cite{kramer2011introduction},

 \begin{equation}
\label{Eq:RVineGeneral}
{\textbf{f}}(P_1,...,P_D)=\prod_{j=1}^{D-1}\prod_{i=1}^{D-j}c_{i,(i+j)}|(i+1),...,(i+j-1)\prod_{k=1}^D {f}_k(P_k) \
\end{equation}
with $c_{i,(i+j)}|i_1,...,i_k$

:= $c_{i,(i+j)}|i_1,...,i_k (F(P_i|P_{i_1},...,P_{i_k}),F(P_j|P_{i_1},...,P_{i_k}))$ for $i,j,i_1,...,i_k$ with $i<j$ and $i_1<...<i_k$.

It is to be noted that the decomposition in \eqref{Eq:RVineDim3} is  not unique and would be different in case of a reordering of the variables indices 1 to $D$.

 The challenge in R-Vine copula set-up is the selection of a specific factorization and  right bivariate copula types,  and the estimation of the copula parameters. AIC, BIC and log-likelihood are the criteria which can be used for pair-copula selection and estimation.
\subsection{Multivariate Scenario Generation}
Once R-Vine structure specifications, pair-copulas types and their parameters are obtained, scenarios can be generated using the inverse probability integral transform. To generate one $D$-dimensional scenario, first, $D$ samples $v_1,...,v_d$ are generated from uniform distribution on $[0,1]$. Then, we set
\begin{equation}
\begin{split}
\label{Eq:Scnarios}
&u_1=v_1 \\
&u_2=F^{-1}(v_2|u_1)\\
&u_3=F^{-1}(v_3|u_1,u_2)\\
&..=....\\
&u_D=F^{-1}(v_D|u_1,...,u_{D-1})
\end{split}
\end{equation}

The formulation  of $F^{-1}(v_i|u_1,...,u_{i-1}$ is given in~\cite{dissmann2013selecting}. 

The samples $u_i\:(\forall i)$ are in unit hypercube space and should be transferred to the original PV power space. Assume  $u_i\:(\forall i)$ are generated for the day $t$, then
\begin{equation}
\label{Eq:ScnariosB}
\hat{P}_{t,d}=\hat{F}^{-1}_{t,d}(u_{t,d}) \quad\forall d
\end{equation}
From \eqref{Eq:ScnariosB}, we get $\hat{\textbf{P}}_{t,d}=\{\hat{P}_{t,1},...,\hat{P}_{t,d}\}$ as a multivariate scenario. Equations \eqref{Eq:Scnarios} and \eqref{Eq:ScnariosB} are repeated $S$ times with different uniform samples to generate $S$ distinctive scenarios.
\subsection{Multivariate Prediction Intervals}
\label{subsection:MPI formulation}
The idea is to find the boundaries for MPI based on the
proportion of the predicted multivariate scenarios which are
enveloped by them. It is desired that if an interval envelops $\alpha$\% of the predicted scenarios, it also encloses the same
proportion of the measured temporal trajectories in a long run. The MPI with nominal coverage rate  $\alpha$\% is obtained as
\begin{enumerate}
	\item Set the boundaries of MPI equal to UPIs with nominal coverage rate $\alpha$\%.
	\item For each lead-time (dimension), increase the width of the MPI by changing its upper and lower limits to the nearest upper and lower scenarios.
	\item Calculate the coverage of MPI by counting the proportion of the scenarios fully inside the boundaries of MPI.
	\item If the coverage is less than $\alpha$\%, repeat steps 2 and 3, otherwise, set the MPIs equal to the intervals found in the previous step.
\end{enumerate}

\subsection{Verification Metrics}
\label{subsection:Verification Metrics}
We will compare the Gaussian  and R-Vine copulas in terms of  their Goodness-of-fit and the quality of their resultant scenarios and MPIs for the case of PV power. 

The corrected Akaike's Information Criterion (AIC) and the Bayesian Information Criterion (BIC) are among the mostly used criteria  for evaluation of the best fitting models\cite{sakamoto1986akaike}. In theory, likelihood is a function of the parameters of a statistical model given observations. It can be calculated as the product of the probability density functions evaluated at the observed data values. Often, it is preferred to work with the natural logarithm of the likelihood function. The so-called log-likelihood function  is defined by
\begin{equation}
\label{Eq: loglok}
\log \mathcal{L}(\theta|P) =\sum _{t=1}^T \log f_t(\theta|P_t)
\end{equation}
with $ \mathcal{L}(\theta|P) $ as the set of model parameter values $ \theta $ given observations $ P $, and $ T $ is the number of observations.

AIC and BIC are defined as
\begin{equation}
\label{Eq: AIC}
AIC=-2\log  \mathcal{L}(\theta|P)+2\kappa
\end{equation}
 The model with the lowest AIC is expected to have a higher goodness-of-fit. 
\begin{equation}
\label{Eq: BIC}
BIC=-2 \log \mathcal{L}(\theta|P)+\kappa\ln(T)
\end{equation}
where $ \kappa $ is the number of estimated parameters in the set $ \theta $.
The model with the lowest AIC and BIC is expected to have a higher goodness-of-fit. 

The skill of multivariate scenarios is verified based on Energy Score (ES) and Variogram-based Score (VS) as the most well-known related elevation metrics~\cite{scheuerer2015variogram}. Both scores are proper and negatively oriented, meaning that a lower score represents a better forecast. Energy score is calculated as
\begin{equation}
\label{Est}
{\rm{E}}{{\rm{S}}_{\rm{t}}} = \frac{1}{S}{\sum\limits_{s = 1}^S {\left\| {{{\rm{\textbf{P}}}_t} - {\rm{\hat {\textbf{P}}}}_t^{(s)}} \right\|}_2} - \frac{1}{{2{S^2}}}{\sum\limits_{s' = 1}^S {\sum\limits_{s = 1}^S {\left\| {{\rm{\hat {\textbf{P}}}}_t^{(s')} - {\rm{\hat {\textbf{P}}}}_t^{(s)}} \right\|}_2 }}
\vspace{0em}
\end{equation} 
\lowercase{with}  $ {\rm{\hat {\textbf{P}}}}_t^{(.)} $ as scenarios distributed according $ \hat{\textbf{F}}_{t} $, and $ \left\| . \right\| $ is the $ D $ dimensional Euclidean norm. 

VS of order $ \gamma $ can be written as   
\begin{equation}
\label{VSt}
V{S_t} \approx {\sum\limits_{i,j = 1}^D {{w_{ij}}\left( {|{P_{t,i}} - {P_{t,j}}{|^\gamma} - \frac{1}{S}\sum\limits_{s = 1}^S {{\rm{|\hat P}}_{t,i}^{(s)} - {\rm{\hat P}}_{t,j}^{(s)}{|^\gamma }}} \right)} ^2}
\end{equation} 
with $w_{ij}$ as non-negative weights.

ES and VS both are averaged over the $ T $ number of forecast time series.

Although during the last decade, a number of verification criteria and skill scores are proposed for evaluation of univariate probabilistic forecasts and multivariate scenario forecasts, the concept of MPI is quite new and the evaluation of MPI is quite challenging at the moment. There is no established framework and formulation for skill verification of MPI. A required feature of any kind probabilistic forecast model, is calibration. It is straightforward to generalize the calibration concept into the multivariate context. Calibration describes how close  the empirical coverage of a MPI to its nominal one is. For MPI, calibration can be calculated by counting the number of measured temporal trajectories which fully lie within the boundaries of each MPI~\cite{bessa2015marginal}. A trajectory of size of
$D$ is enclosed by an interval if for $d = 1,...,D$, the
trajectory is inside or equal to the limits of that MPI.  The ratio of enclosed trajectories to the total number trajectories is known as the empirical coverage. In this study, the sharpness of MPIs is decided based on their volumes where a lower volume shows a better sharpness.
% For a MPI with $\alpha$ nominal probability, and as lower and upper limits, the empirical coverage ($\hat{\alpha}_i$) is calculated as
% \begin{equation}
%\label{Eq:Coverage}
%\vspace{0em}
%\hat{\alpha}_i = \frac{1}{N} \sum_{t=1}^{N} {\xi_{t,k}^{\alpha_i}}\\
%\end{equation}

The volume of a MPI is calculated as
\begin{equation}
	\label{Eq:Volume_MPI}
V_t=\prod_d(h_{d,t}-l_{d,t}) \quad\forall t
\end{equation}
with $V_t$ as the volume of MPI at time $t$, $h_{d,t}$ and $l_{d,t}$ as the upper and lower
bounds of the MPI for  $t^{th}$ trajectory and dimension $d$, respectively.
\section{Results}
\begin{figure*}[t!]
	
	\begin{tabular}[c]{ccc}
		\begin{subfigure}{.33\textwidth}
			\centering
			\includegraphics[width=.9\linewidth]{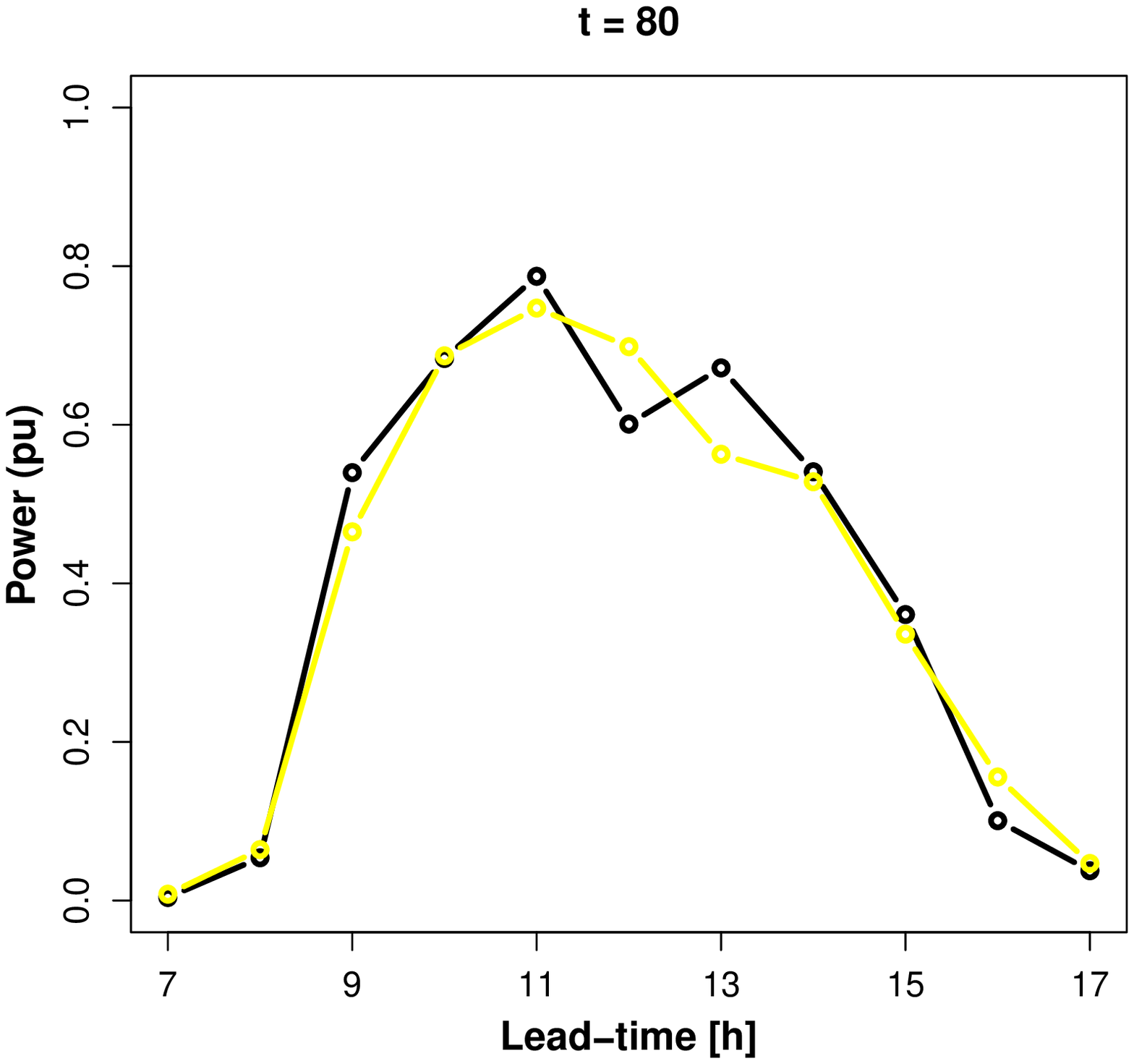}
			\caption{}
			\label{fig:UPI}
		\end{subfigure}
		\begin{subfigure}{.33\textwidth}
			\centering
			\includegraphics[width=.9\linewidth]{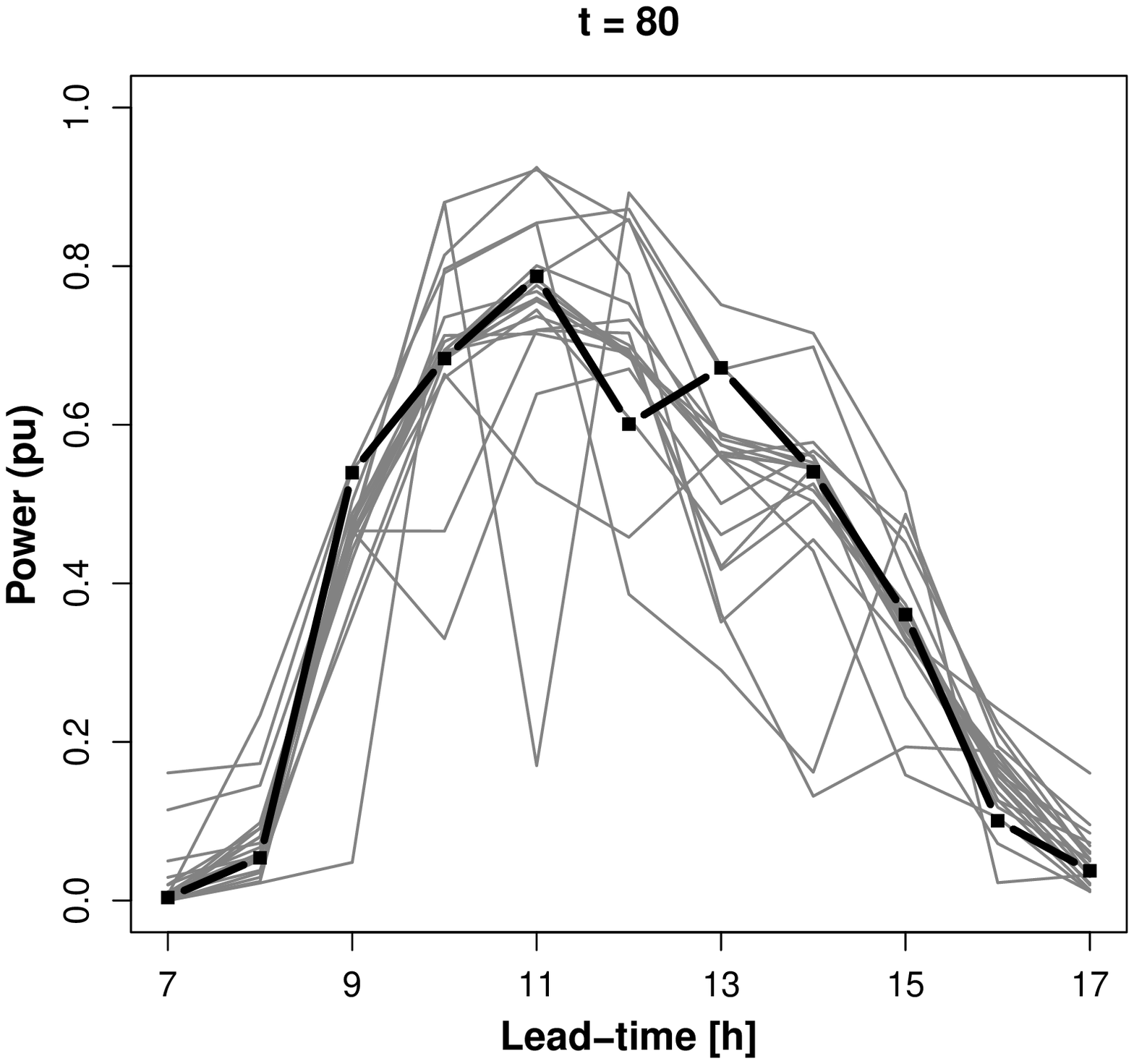}
			\caption{}
			\label{fig:Scenarios1}
		\end{subfigure}
		\begin{subfigure}{.33\textwidth}
			\centering
			\includegraphics[width=.9\linewidth]{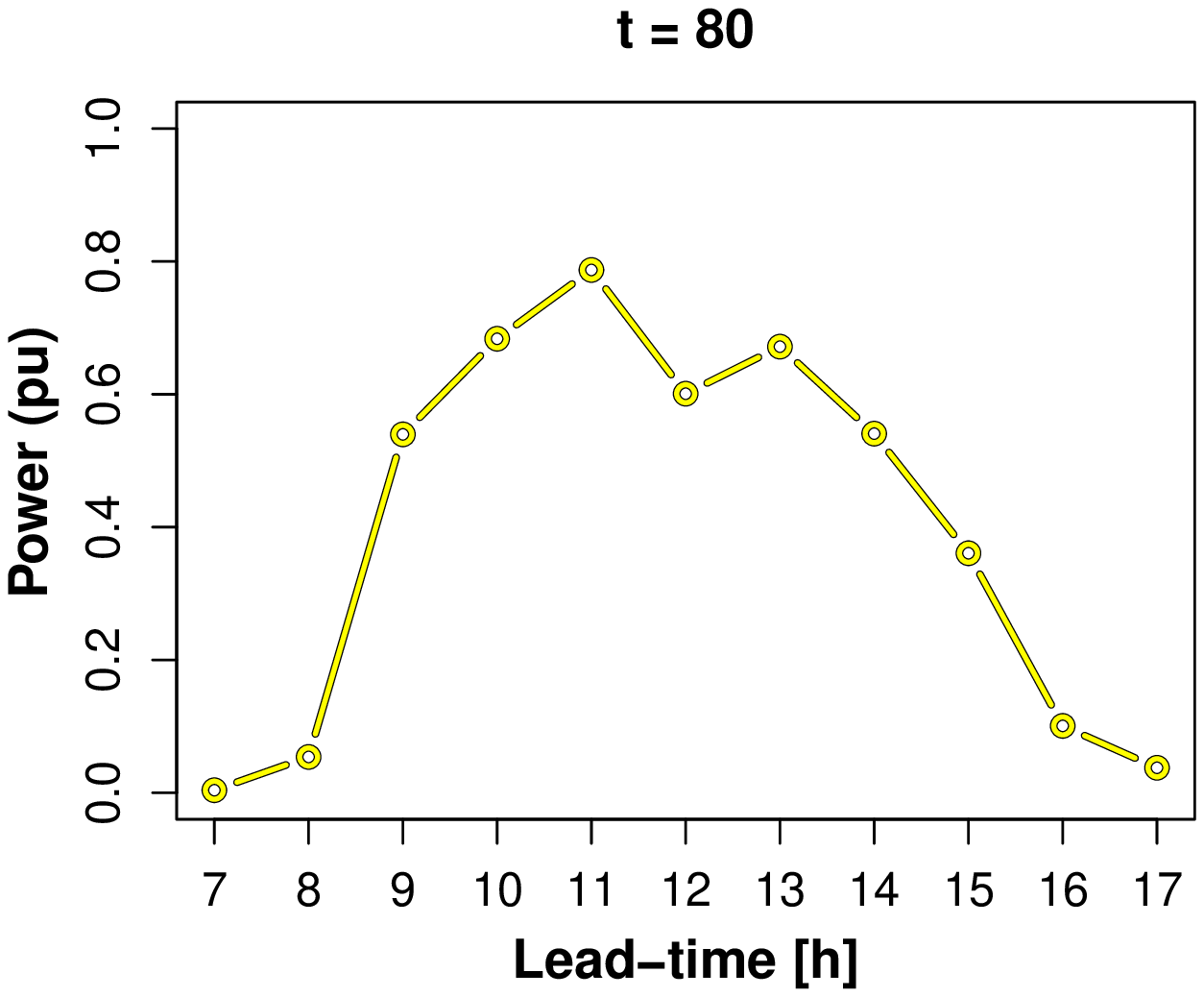}
			\caption{}
			\label{fig:MPIIntervals}
		\end{subfigure}
	\end{tabular}
	\caption{(a) PV observations (yellow colour curves) along with 19 univariate prediction intervals with coverage ranging from 0.05 to 0.95 by 0.05 increments (from the darkest to the lightest, (b) PV observations (dark black colour curves) along with 20 generated space-time trajectories (grey colour curves), (c) PV power observations (yellow colour curves)  along with 19 MPIs with nominal coverage ranging from 0.05 to 0.95 by 0.05 increments (from the darkest to the lightest)}
	\label{fig:UPISCMPI}
	\vspace{-0.5em}
\end{figure*}
The dataset used in this work for investigations includes 12  independent  variables as the output of NWPs provided by ECMWF as explanatory variables. Data covers the period from April 2012 to the end of June 2014. Real measurements and NWPs for  three adjacent PV  zones are available.  The first zone is studied here. Data for the  period from April 2012 to the end of May 2013 is considered as training subset to train quantile regression model. The temporal resolution of data is of one hour. The evaluation subset covers the data from June 2013 to the end of January 2014.  Power measurements are normalized by the nominal capacity of the corresponding PV installation. All the analysis provided below are based on the results obtained for the evaluation subset. AIC is considered as criterion for pair-copula selection and estimation. For VS, $w_{ij} \: \forall i,j$ is considered to be equal to 1 and $\gamma$ is set to 0.5~\cite{golestaneh2016generation}.

Day hours from 7 am to 5 pm are covered. Therefore, the multivariate predictive distributions and multivariate scenarios are of dimension 11.  UPI refers to the prediction intervals generated by quantile regression model for each lead-time individually without considering the correlations between various lead-times. Multivariate scenarios are generated for both Gaussian  and R-Vine copulas. Then, the  scenarios are deployed to obtain the MPIs. The number of scenarios is considered to be 500. UPIs and MPIs with nominal coverages ranging from 0.05 to 0.95 by 0.05 are generated and examined.

Univariate quantiles are generated using quantile regression for each lead-time individually. Once predictive quantile are obtained, central UPI are a natural by-product allowing for better visualization of forecast uncertainties~\cite{golestaneh2016very}.

Moving from univariate modeling to multivariate structure,  multivariate predictive distributions are generated based on R-Vine and Gaussian copulas as explained in subsection \ref{subsection:Vine}. 
 
Multivariate scenarios are obtained by randomly sampling from multivariate predictive densities. Following that, MPI are generated based  on the adjusted interval method explained in subsection \ref{subsection:Vine}. 
 
Fig. \ref{fig:UPISCMPI} illustrates UPIs, multivariate scenarios and MPIs for a randomly selected day from the evaluation data. 
 
Comparing Fig. \ref{fig:UPI} and Fig.  \ref{fig:MPIIntervals} reveals that MPI are far wider than UPIs because they are supposed to describe the uncertainty in all successive lead-times simultaneously. This is more pronounced in MPIs with lower nominal coverages. As one can see in  Fig.  \ref{fig:MPIIntervals}, the differences between widths of the intervals with nominal coverage lower and higher than 50\%  are very low.

\begin{figure}[t!]
	\centering
	\includegraphics[width=0.85\linewidth]{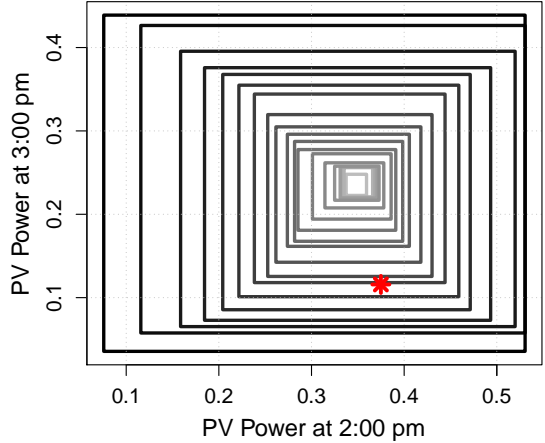}
	\caption{Nineteen  MPIs with probabilities ranging from 0.05 to 0.95 by 0.05 increments (from the lightest to the darkest), for a randomly selected day from the evaluation data. The red point shows the measured PV power for the same hours and date.\label{fig:Boxes}}
	\vspace{-0.5em}
\end{figure}

To get a better idea of the uncertainty regions introduced by the MPIs, bivariate MPIs describing correlated PV power generations at 2:00 pm and 3:00 pm are generated. Bivariate MPIs for a randomly selected day are shown in \ref{fig:Boxes}.  The red point shows the measured PV power for the same time and date. As can be seen in the figure, the MPI corresponds to a higher probability is larger than the one with a lower nominal coverage. In order to account for correlation of intermittent generation, the  regions similar to those shown in Fig. \ref{fig:Boxes} should be utilized as the constraints in the operational problems such as optimal power flow and unit commitment.

\begin{figure}[t!]
	\centering
	\includegraphics[width=0.85\linewidth]{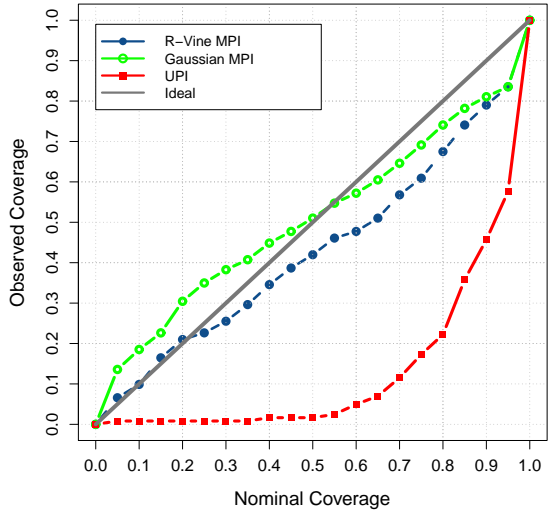}
	\caption{Calibration of the intervals with nominal coverage ranging from 0.05 to 0.95 by 0.05 increments. \label{fig:QQMPI}}
	\vspace{-0.5em}
\end{figure}

Fig. \ref{fig:QQMPI} illustrates the reliability (calibration) digram for UPI along with Gaussian, R-Vine MPIs.  In the reliability digram the observed (empirical) coverage of probabilistic forecasts are shown against their corresponding nominal coverages. Therefore, these diagrams inform about  how well the predicted MPIs of an event correspond to their observed frequencies. In Fig. \ref{fig:QQMPI}, the gray solid line represents perfect calibration and the curves for the MPIs should be as close as possible to that of the ideal case. The empirical coverages are calculated as explained in subsection \ref{subsection:Verification Metrics}. One can observe that in Fig. \ref{fig:QQMPI}, UPI shows very low calibration.  UPIs are generated for each lead-time individually and independently. Therefore, they do not carry information about the interdependence of PV generation in successive lea-times. When they are examined on how well they can envelop measured temporal trajectories, they are expected to show low calibration. Gaussian MPIs tend to overestimate uncertainty for low nominal probabilities and underestimate it for higher nominal probabilities. This is different with R-Vine MPIs which show mostly an underestimation trend.
\begin{table*}[h]
	\centering
	\caption{Comparative results in terms of Goodness-of-fit tests and multivariate trajectory scores}
	\label{Table:BIC}
	\centering
	\resizebox{\textwidth}{!}{%
		%	\begin{adjustbox}{max width=\textwidth}
		\begin{tabular}{c|ccccccc}
			\specialrule{1.5pt}{0pt}{0pt}
			& \textbf{Log-likelihood} & \textbf{AIC} & \textbf{BIC} & \textbf{ES} & \textbf{VS} &\textbf{Average Deviation (\%)} & \textbf{ Average Volume of 95\% MPIs}\\ \specialrule{1pt}{0pt}{0pt}
			\textbf{R-Vine Copula} & 616.939           & -1081.879         & -830.828           & 6.997$\times 10^{-2}$     &  91.361$\times 10^{-2}$    & 7.605 & 4.324 $\times 10^{-3}$ \\ \hline
			\textbf{Gaussian Copula} & 396.573           & -683.145           & -460.281           & 7.008$\times 10^{-2}$       & 91.637$\times 10^{-2}$   & 6.303 & 4.894 $\times 10^{-3}$  \\ \specialrule{1.5pt}{0pt}{0pt}
		\end{tabular}
	}
\end{table*}

Table \ref{Table:BIC} summarizes the comparative results for Gaussian and R-vine based approaches in terms of various evaluation metrics. Looking at goodness-of-fit metrics, the multivariate predictive density set-up based on R-Vine copula shows much higher goodness-of-fit comparing to the one formed by Gaussian copula. The R-Vine predictive density surpasses the Gaussian density in terms of BIC, AIC and log-likelihood. However, the interesting observation is that, the higher goodness-of-fit does not necessarily leads to MPI or multivariate scenarios with higher predictive performance.  The scenarios generated by sampling from R-Vine density still show better energy and variogram-based scores but the betterment is not noticeable as it is for goodness-of-fit metrics. The performance of MPIs are compared in terms of their average deviations from their nominal coverages. The deviations are averaged over 19 MPIs with coverage ranging from 0.05 to 0.95 by 0.05 increments. The average volume of MPIs with nominal coverage 95\% generated for all days in the evaluation set is considered as representative of sharpness. As given in  Table \ref{Table:BIC}, R-Vine predictive densities offer slightly better sharpness, while Gaussian predictive densities present slightly higher calibration. Without having any established skill score to provide a unique score  accounting for both reliability and sharpness together, it is difficult to judge which method has a better predictive performance to generate MPIs of PV power.  This would be the focus of future works to explore and develop proper skill scores for evaluation of MPIs.

\section{Conclusion}
Univariate prediction intervals are produced here for PV power using quantile regression. The quantiles are used as inputs to model R-Vine and Gaussian multivariate predictive densities.  Log-likelihood, BIC and AIC are considered here as  goodness-of-fits metrics. The results suggest a much higher fitness of R-vine copula over Gaussian one in characterizing  PV power at correlated  successive leas-times. However, the skill score values calculated for multivariate scenarios generated from these densities show that R-Vine and Gaussian scenarios present very close quality in terms of  energy  and variogram-based scores. The scenarios then are considered as the building blocks to generate multivariate intervals. The results show that both Gaussian and R-Vine copulas lead to MPIs with reasonable calibration and sharpness. The MPIs can serve as very critical inputs in decision-making problems in power systems such as unit-commitment and optimal power flow which need information about how random variables vary over successive lead-times.

\vspace{-0.3em}

%\section*{Acknowledgment}
\bibliographystyle{IEEEtran}  

\bibliography{reftest}

\end{document}